\documentclass[12pt]{article}

\usepackage{indentfirst} 
\usepackage[T2A]{fontenc}
\usepackage[cp1251]{inputenc}
\usepackage[english]{babel}
\usepackage{amssymb,amsfonts,amsmath,mathtext,cite,enumerate,float}
\usepackage[dvips]{graphicx}

\makeatletter

\renewcommand{\@biblabel}[1]{#1.\hfill}
\makeatother

\usepackage{geometry} 
\title{Stability of Hartmann flow with the convective approximation}
\author{I.Yu.~Kalashnikov}

\begin{document}

\begin{flushright}
{\tiny Published in the Magnetohydrodynamics. Vol. 50 (2014), No. 4, pp. 353–359.}
\end{flushright}

\begin{center}
{\LARGE Stability of Hartmann flow with the convective approximation}
\end{center}

\begin{center}
{I.Yu.~Kalashnikov}
\end{center}

\begin{center}
National Research Nuclear University MEPhI \\ Schmidt Institute of Physics of the Earth \\  Moscow, Russia.
\end{center}
\begin{center}
\textit{kalasxel@gmail.com}
\end{center}

\begin{abstract}
 This report focuses on the linear analysis of a plane-parallel flow stability in transverse magnetic field (Hartmann flow) within the convective approximation. We obtain and solve equations describing the perturbation growth. We found the perturbation modes and their non-excitation conditions. We obtain the equation for the instability increment and show that it has an instable root. Also we shown that resulting instabilities qualitatively agree with the experimental data.
\end{abstract}

\section*{Introduction}

\begin{figure}
\centering
\includegraphics[width=7cm]{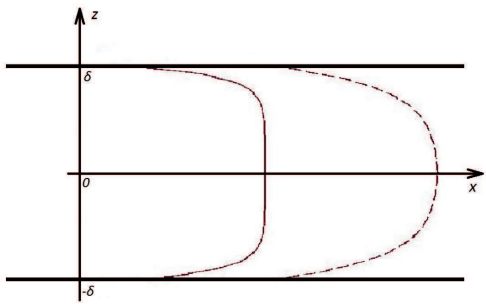}
\caption{Profiles of the velocity field for the Poiseuille flow (dash line) and Hartmann flow (solid line).}
\label{gart}
\end{figure}

Hartmann flow is a steady stream between two fixed infinite parallel planes arising due to the pressure drop that occurs in a magnetic field directed orthogonally to planes. We choose the $z$ axis to be co-directional with the external magnetic field $B_0$, and the $x$ axis direct along the stream (see Fig. \ref{gart}). For such a flow there is an exact solution:
\begin{equation}
\label{garV}
 V_x(z) = \frac{k_2 \delta}{k_1 \sinh(k_1 \delta)} (\cosh(k_1 \delta) - \cosh(k_2 z)),
\end{equation}
\begin{equation}
\label{garB}
 \sqrt{\frac{\nu_m}{4 \pi \rho \nu}} B_x = -\frac{k_2}{k_1} z + \frac{k_2 \delta}{k_1 \sinh(k_1 \delta)} \sinh(k_1 z),
\end{equation}
where $k_1 = B_0 / \sqrt{4\pi \rho \nu \nu_m}$, $k_2 = -(1/\rho\nu) (\partial p / \partial x)$. Constants $\nu$ and $\nu_m$ are kinematic and magnetic viscosity respectively, $\rho$ is a density of the fluid.

From (\ref{garV}), (\ref{garB}) we see that with increasing transverse magnetic fields velocity profile becomes flatter. This is due to the Lorentz force acting on different areas differently. Where the speed of the current is less than the average speed, the electric current flows along the $y$ axis and against it when the speed exceeds the average value. In the first case, the Lorentz force accelerates the flow and in the second case it slows the flow down. This leads to a flattening of the velocity profile. This flatness is characterized by the Hartmann number: $Ha = B_0 / \sqrt{4 \pi \rho \nu \nu_m}$.

Stability of the Hartmann flow was first considered in \cite{Lock} and there was obtained a dimensionless equation similar to the hydrodynamic equations of Orr – Sommerfeld:

\begin{equation}
 (u-c)(\psi'' - \alpha^2 \psi) - u'' \psi + \frac{i}{\alpha Re}(\psi^{(4)} - 2 \alpha^2 \psi'' +\alpha^4 \psi) = \frac{i Ha^2}{\alpha Re} \psi'' ,
\end{equation}
where $u$ means an unperturbed velocity, $\psi$ - stream function perturbation, $\alpha$ and $c$ - dimensionless wave frequency and phase velocity of the perturbation. The boundary conditions are $\psi(\pm 1) = \psi'(\pm 1) = 0$. The right side of the equation which takes into account the direct effect of the magnetic field on the perturbation is negligible. So we obtain the usual equation of Orr - Sommerfeld, but with Gartmon velocity profile,
which can be replaced by a close one: $u=1-exp(-Ha (1+|z|))$.
Thus the influence of the magnetic field is taken into account only by changing the velocity field
and we obtain a (quite expected) result that
with an increase of the magnetic field stability increases too. For sufficiently large ($Ha> 20$) magnetic field there is a linear dependence of critical Reynolds number on the Hartmann number: $Re_c = 5 \cdot 10^4 Ga$.


In \cite{Veli} the stability of the ideal conducting fluid between two coaxial rotating cylinders is considered. They write down equations describing an evolution of perturbations and by analyzing them (but not solving) derive a  stability criterion of the current in the layer between cylinders (Velikhov criterion) $\partial \Omega(r) / \partial r \geq 0 $, where $\Omega$ is the angular velocity of the fluid layer. Comparing this criterion with the hydrodynamic Rayleigh criterion $\partial (r^2 \Omega)/\partial r \geq 0$ we can conclude that the magnetic field destabilizes the flow in a rotating cylinder up to a certain value of the magnetic field, and strong magnetic field stabilizes the flow. This is due to the fact that in a weak magnetic field electrodynamic forces are already affecting the nature of small-scale motions, and the effect of freezing has not yet manifested.

Also worth noting the work of \cite{More} in which the transition to turbulence due to the instability Hartmann layer and conditions of turbulence suppression (laminarization) have been studied experimentally . It has been found that when the parameter $R=Re/Ha > 380$ the flow becomes turbulent. Numerical simulations \cite{Krasnov} give approximately the same result. Because $R$ is inversely proportional to the magnetic field it may be concluded that a weak magnetic field destabilizes the current.

The question then arises: how important is the role of rotation in the formation of magnetorotational instability? Is there a parallel flow instability and under what conditions they are excited? Let’s try to answer these questions.

\section*{Two-dimensional perturbations}
Assume instabilities are convective, i.e. perturbations that arise at any point do not have enough time to develop and are carried over beyond the real pipe. But since the magnetic field has (because of embeddedness) an inhibitory effect, for a feasibility of such an assumption the magnetic field should be small.


The magnetic field can be compared to the magnetic viscosity by choosing multipliers to make the dimensions to match. The same result can be achieved by requiring the Alfven speed  $B_0 / \sqrt{4\pi\rho}$ to be much smaller than some characteristic velocity of the fluid. We can construct this value from dimensional parameters of the liquid in three ways: $ \nu_m / \delta$, $ \nu / \delta $ and $ \sqrt{\nu \nu_m} / \delta $. Since the embeddedness is affected only by the magnetic viscosity, it is logical to choose the first way. Thus we obtain:

 \begin{equation}
\label{vmor}
  \frac{B_0 \delta}{\sqrt{4\pi \rho}\nu_m} \ll 1 .
\end{equation}


In this approximation we suppose that the perturbation does not evolve and moves along the main flow, ie  $ \partial / \partial x = 0 $. Then we investigate stability of the system of equations:

\begin{equation}
\frac{\partial \textbf{V}}{\partial t} + ( \textbf{V} , \nabla ) \textbf{V} = \frac{1}{4 \pi \rho} ( \textbf{B} , \nabla ) \textbf{B} - \frac{1}{\rho} \nabla( P + \frac{ {\textbf{B}}^2}{8 \pi} ) + \nu \nabla^2 \textbf{V},
\end{equation}

\begin{equation}
\label{induk}
\frac{\partial \textbf{B}}{\partial t} = \nabla\times(\textbf{V}\times \textbf{B}) + \nu_m \nabla^2 \textbf{B},
\end{equation}

\begin{equation}
\label{divV}
  \nabla\cdot \textbf{V} = 0,
\end{equation}

\begin{equation}
\label{divB}
  \nabla\cdot \textbf{B} = 0,
\end{equation}
 assuming that the main flow obeys equations \eqref{garV}, \eqref{garB} and all instabilities are convective.

We make transformations $ \textbf{B} \rightarrow \textbf{B} + \textbf{b} $, $ \textbf{V} \rightarrow \textbf{V} + \textbf{v} $, $ P \rightarrow P + \varphi $, where $ \textbf{V} = ( V(z) , 0 , 0 ) $, where $ \textbf{B} = ( B(z) , 0 , B_0 ) $ is known. Then we leave only linear terms of perturbations. Then, since the movement is infinite in time $t$ and coordinate $y$, we assume that the perturbation is periodic in terms of these variables: $ f(y,z,t) \rightarrow f(z) exp(i \gamma t - i k y) $. I.e. in each layer $d z$ there propagates a plane wave. The boundary conditions for perturbations correspond to adhesion and impermeability conditions; the wall is non-magnetic, so perturbations on them tend to zero:

 \begin{equation}
  \textbf{b}(\pm\delta) = \textbf{v}(\pm\delta) = \varphi(\pm\delta) = 0.
\end{equation}

Since there is no field sources, it is possible to introduce the vector potential, and as $ \partial / \partial x = 0 $, then $y$ and $z$ rotor components comprise only one term. Denoting then $b_x=b$ and $v_x = v$ we write:

\begin{equation}
  \textbf{b}=(b, \frac{\partial a}{\partial z} , - \frac{\partial a}{ \partial y} ) = (b, \frac{\partial a}{\partial z} , i k a ),
\end{equation}

\begin{equation}
  \textbf{v}=(v, \frac{\partial q}{\partial z} , - \frac{\partial q}{ \partial y} ) = (v, \frac{\partial q}{\partial z} , i k q ).
\end{equation}


Due to these transformations equations (\ref{divV}) and (\ref{divB}) disappear, and $y$ and $z$ components of the equation (\ref{induk}) are identical. Then the system of equations describing the amplitude of a plane wave perturbation has the form:

\begin{equation}
 i \gamma v + i k V' q - \frac{B_0}{4 \pi \rho} \frac{d b}{d z} - i k \frac{B'}{4 \pi \rho} a +\nu k^2 v -\nu \frac{d^2 v}{d z^2} = 0,
\end{equation}

\begin{equation}
 i \gamma b - i k V' a - B_0 \frac{d v}{d z} + i k B' q +\nu_m k^2 b -\nu_m \frac{d^2 b}{d z^2} = 0,
\end{equation}

\begin{equation}
 \frac{d M}{d z} - i k N = \frac{B_0}{4 \pi \rho} \nabla^2 a,
\end{equation}

\begin{equation}
 \frac{d N}{d z} + i k M = 0,
\end{equation}
where stand out two quantities:

\begin{equation}
\label{M}
 M = i \gamma q +\nu k^2 q - \nu \frac{d^2 q}{d z^2},
\end{equation}

\begin{equation}
\label{N}
 N = \frac{\varphi}{\rho} + \frac{B b}{4 \pi \rho}.
\end{equation}


For $N$ the boundary conditions are obviously zero: $ N(\pm \delta) =0 $. However, another condition is needed to determine the possible wave numbers $k$. Note that if we make a replacement in the expression for M: $z \rightarrow -z$ (or $ k \rightarrow -k $) it does not change. So we can say that $ M(\delta) = M(-\delta) $. This statement can also be proven from similarity of the expression for $M$ with the heat equation (in $x$ - space). Then $M$ acts as a heat source. And since the walls are identical, they will generate perturbations in the same manner.

Then from (\ref{M}), (\ref{N}) and the boundary conditions which are set above we obtain the eigenvalues of the wave number:

\begin{equation}
 k = -i \frac{\pi}{\delta} n, \;\;\; n\in \mathbb{Z}.
\end{equation}


I.e. for $k \neq 0$ there is an instability increasing either to the right or left relative to the flow. 
Nonzero modes will not be excited if $\delta$ is sufficiently large. We can compare $\delta$ with the parameters of liquid in three ways, but since this instability is due to hydrodynamic and electrodynamic forces then we choose the option for $\delta$ where $\nu$ and $\nu_m$ are included equally. Therefore, considering the convection assumption we have a range for $\delta$:

\begin{equation}
\label{delta}
 \frac{\sqrt{4\pi\rho \nu \nu_m}}{B_0} \ll \delta \ll \frac{\sqrt{4\pi\rho} \nu_m}{B_0},
\end{equation}
which implies that $\nu \ll \nu_m $ or magnetic Prandtl number $Pr_m \ll 1$ and also $Ha \gg 1$.

\section*{One-dimensional perturbations.}


It should be noted that for $k\neq 0$ the obtained system can be solved exactly, but we  restrict our investigation to a one-dimensional flow in the $y$-stable region. Then vector perturbations are two-dimensional (no $z$ component). We obtain two independent systems for potentials  (\ref{kompyPOTa}), (\ref{kompyPOTq}) and for the components of perturbations (\ref{kompyB}), (\ref{kompyV}):

\begin{equation}
\label{kompyPOTa}
    i \gamma a - B_0 \frac{d q}{d z} - \nu_m \frac{d^2 a}{d z^2} = 0,
\end{equation}

\begin{equation}
\label{kompyPOTq}
    i \gamma q - \frac{B_0}{4 \pi \rho} \frac{d a}{d z} - \nu \frac{d^2 q}{d z^2} = 0,
\end{equation}

\begin{equation}
\label{kompyB}
    i \gamma b - B_0 \frac{d v}{d z} - \nu_m \frac{d^2 b}{d z^2} = 0 ,
\end{equation}

\begin{equation}
\label{kompyV}
    i \gamma v - \frac{B_0}{4 \pi \rho} \frac{d b}{d z} - \nu \frac{d^2 v}{d z^2} = 0,
\end{equation}
the pressure disturbance is expressed through disturbance of the magnetic field as follows:

\begin{equation}
 \varphi = - \frac{ B b }{4\pi}.
\end{equation}
Boundary conditions are as follows:
\begin{equation}
 b(\pm\delta) = v(\pm\delta) = \frac{d a}{d z} (\pm\delta) = \frac{d q}{d z} (\pm\delta) = 0.
\end{equation}


Note that when $\nu = \nu_m$ there is a symmetry in both systems: after a replacement $b \rightarrow v \sqrt{4\pi\rho}$ (or $a \rightarrow q \sqrt{4\pi\rho}$) the form of both equations doesn't change. Therefore, the resulting spectrum is degenerate, so the eigenfunctions can be found in the form $b= \sigma v \sqrt{4\pi\rho}$, where $\sigma = \pm 1$. We obtain the eigenvalues corresponding to a stable flow:

\begin{equation}
 i\gamma = - \frac{\nu}{(2\delta)^2} ( \pi^2 n^2 + {Ha}^2) \leq 0, \,\,\,  n\in \mathbb{Z}.
\end{equation}


This means that for the close values of kinematic and magnetic viscosity small one-dimensional perturbations are damped and the flow is stable. However, as follows from (\ref{delta}), there already exist some undamped perturbation modes.


Solution of the system (\ref{kompyPOTa}), (\ref{kompyPOTq}) and (\ref{kompyB}), (\ref{kompyV}) gives us the same eigenvalues $\lambda$ and leads to the same equation for the eigenvalues of the increment $\gamma$. We solve (\ref{kompyB}), (\ref{kompyV}) by seeking a solution in the form $b = b_0 e^{\lambda z}$, $v = v_0 e^{\lambda z}$. It leads to the eigenvalues:

\begin{equation}
 {\lambda}^2 = \frac{1}{2} \left [ i \gamma \left ( \frac{1}{\nu} + \frac{1}{\nu_m} \right ) + \frac{{B_0}^2}{4 \pi \rho \nu \nu_m} \pm \sqrt{ \left [ i \gamma \left ( \frac{1}{\nu} + \frac{1}{\nu_m} \right ) + \frac{{B_0}^2}{4 \pi \rho \nu \nu_m} \right ]^2 - \frac{4 \gamma^2}{\nu \nu_m} } \right ].
\end{equation}
Then we have a solution in the form:
\begin{equation}
\label{Beq}
    b(z)=b_1(C_1 e^{\lambda_1 z} + C_2 e^{-\lambda_1 z}) + b_2(C_3 e^{\lambda_2 z} + C_4 e^{-\lambda_2 z} ),
\end{equation}

\begin{equation}
\label{Veq}
    v(z)=v_1(C_1 e^{\lambda_1 z} - C_2 e^{-\lambda_1 z}) + v_2(C_3 e^{\lambda_2 z} - C_4 e^{-\lambda_2 z} ),
\end{equation}
where the eigenvectors are chosen in the following way: $b_1 = B_0 \lambda_1 \delta $, $b_2 = B_0 \lambda_2 \delta$, $v_1 = \delta ( i \gamma - \nu_m {\lambda_1}^2)$, $v_2 = \delta ( i \gamma - \nu_m {\lambda_2}^2)$ ($\lambda_1$ taken with the positive sign before the square root). With such a choice the coefficients $C$ are dimensionless.

Due to the boundary conditions we obtain a homogeneous system of four linear algebraic equations, which admits nontrivial solutions for $\lambda$ satisfying the equation:
\begin{equation}
\label{chsh}
 b_1 v_2 \cosh(\lambda_1 \delta) \sinh(\lambda_2 \delta) = v_1 b_2 \sinh(\lambda_1 \delta) \cosh(\lambda_2 \delta).
\end{equation}
Equation (\ref{chsh}) can be factorized:
\begin{equation}
\label{sis1}
    b_1 v_2 = B_0 V_0 \tanh(\lambda_1 \delta),
\end{equation}

\begin{equation}
\label{sis2}
    v_1 b_2 = B_0 V_0 \tanh(\lambda_2 \delta),
\end{equation}

\begin{equation}
\label{chch}
  \cosh(\lambda_1 \delta) \cosh(\lambda_2 \delta) = 0,
\end{equation}
where $V_0$ is a quantity having the dimension of velocity. 
Choose for example $V_0 = \nu/\delta$, then $\gamma$ can be easily made dimensionless $\gamma\delta^2 / \nu \rightarrow \gamma $. Eigenvalues $\lambda$ must satisfy either both equations (\ref{sis1}), (\ref{sis2}) or equation (\ref{chch}).

\begin{figure}
\centering
\includegraphics[width=9cm]{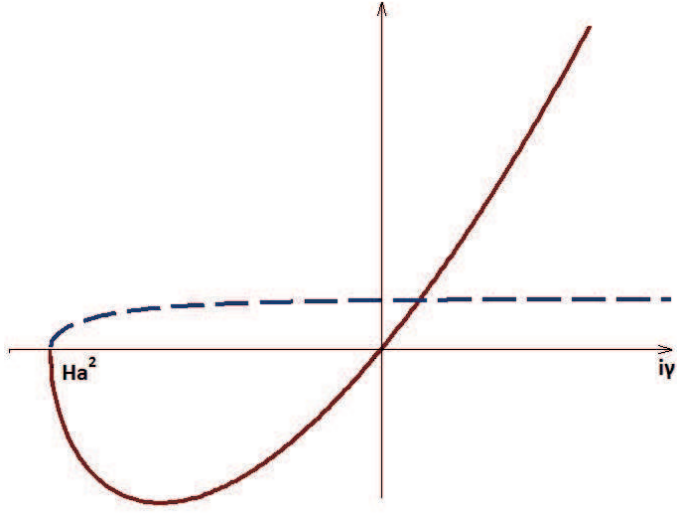}
\caption{The real roots of the equation (\ref{transPR}). Solid line: $i\gamma \sqrt{i\gamma + {Ha}^2}$. Dashed line: $\tanh \sqrt{i\gamma + {Ha}^2}.$ }
\label{korni}
\end{figure}

Because we restricted ourselves to the area where $Pr_m \ll 1$ we can make a Taylor expansion in terms of this parameter $\lambda$. As a first approximation we set $Pr_m = 0$. Then we obtain the following eigenvalues: $\lambda_1 \delta \simeq \sqrt{ i \gamma + {Ha}^2}$, $\lambda_2 \simeq  0$. Equation (\ref{chch}) gives stable roots:
\begin{equation}
 i\gamma = - Ha^2 - (\frac{\pi}{2} + \pi n)^2 < 0, \;\;\; n\in \mathbb{Z}.
\end{equation}
The system (\ref{sis1}), (\ref{sis2}) becomes a single equation, because the latter is satisfied identically:
\begin{equation}
\label{transPR}
 i\gamma \sqrt{i\gamma + {Ha}^2} = \tanh \sqrt{i\gamma + {Ha}^2}.
\end{equation}

One of the roots (stable one) is immediately visible: $ i\gamma = - {Ha}^2 < 0 $. Another root corresponds a pure imaginary increment indicating the flow instability: $i\gamma > 0$ (see Fig. \ref{korni}). Numerical solution of the equation (\ref{transPR}) gives us other roots among which there are those corresponding to the instability.

\section*{Discussion and conclusions}
\begin{figure}[h]
\centering
\includegraphics[width=8cm]{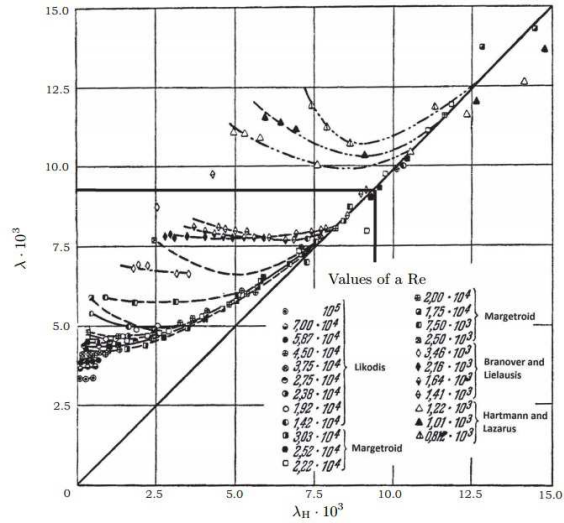}
\caption{Experimental data on the resistance coefficient in comparison with the theory for Hartmann flow. Highlighted region denotes the range of applicability of the convective approximation.}
\label{opit}
\end{figure}

Fig. \ref{opit} shows a summary of the experimental results on the study of the stability of the Hartmann flow \cite{Bran}. In these experiments the resistance coefficient $\lambda=-2p'\delta/\rho V^2$ has been measured. For the Hartmann flow it has the form: $\lambda_H \simeq 2 Ha / Re$. At the figure the deviation from a bisector of the coordinate angle means that the flow for the given parameters is already turbulent.

From (\ref{vmor}) follows that  $Ha\sqrt{Pr_m}\ll 1$, so in the convective approximation $\lambda \ll Ha^{-1} {Re_m}^{-1}$. Since the experiments were carried out at $Ha \sim 10^3$ and $Re_m \sim 10^{-1}$ (mercury) it is possible to say that $\lambda_H \ll 10^{-2}$. And as can be seen from Fig. \ref{opit} in this field the flow is not laminar. Thus the found instability really takes place.

Fulfillment of the condition (\ref{delta}) provides only the absence of perturbations aimed parallel to the planes (i.e. the absence of $x$ and $y$ components of perturbation). As can be seen from (\ref{transPR}) with an increase of $ Ha $ the instability becomes suppressed. This can be explained by the fact that with increasing magnetic field (and as a consequence an increase in the Hartmann number) embeddedness effect begins to dominate, making it difficult to form instabilities in the initially laminar flow. But the equation (\ref{transPR}) is obtained in the limit $ Pr_m \rightarrow 0 $. To determine more precise conditions under which the flow could be sustained one needs a detailed analysis of (\ref{chsh}).

It should also be noted that in the limit $ Ha \rightarrow 0 $ the instability is still present, although the flow itself becomes a Poiseuille flow. Absence of such a limit transition was noticed in \cite{Veli} and there this is explained by the fact that a non-zero magnetic field already generates instabilities which then can develop without the participation of the magnetic field.

\subsection*{Acknowledgements} The author expresses his gratitude to D. Sokolov for advice during the solving, Yu. Kolesnikov, P. Frick and V. Chechetkin, who provided valuable comments, as well as Yu. Ivanova and C. Maslov for their assistance in preparing this article.

The author was funded from RFBR research grant No. 14-29-06086.

\newcommand{\noopsort}[1]{}

\end{document}